\documentstyle[prc,aps,psfig]{revtex}
\begin{document}
\twocolumn
 
\title{First observation of $^{54}$Zn and its decay by two-proton emission}

\author{B. Blank$^1$, A. Bey$^1$, G. Canchel$^1$, C. Dossat$^1$, A. Fleury$^1$, J. Giovinazzo$^1$, I. Matea$^1$, N. Adimi$^2$, 
F. De Oliveira$^3$, I.~Stefan$^3$, G. Georgiev$^3$, S. Gr\'evy$^3$, J.C. Thomas$^3$, C. Borcea$^4$, D. Cortina$^5$, M. Caamano$^5$, 
M. Stanoiu$^6$, F.~Aksouh$^7$, B.A. Brown$^8$, F.C. Barker$^9$, W.A. Richter$^{10}$}

\address{
$^1$ CENBG, Le Haut Vigneau, F-33175 Gradignan Cedex, France \\
$^2$ Facult\'e de Physique, USTHB, BP32, El Alia, 16111 Bab Ezzouar, Alger, Algeria \\
$^3$ Grand Acc\'el\'erateur National d'Ions Lourds, B.P. 5027, F-14076 Caen Cedex, France \\
$^4$ Institute of Atomic Physics, P.O. Box MG6, Bucharest-Margurele, Romania \\
$^5$ Dpto. de Fisica de Particulas, Universidad de Santiago de Compostela, E-15782 Santiago de Compostela, Spain \\
$^6$ Institut de physique nucl\'eaire d'Orsay, 15 rue Georges Clemenceau, F-91406 Orsay Cedex, France \\
$^7$ Instituut voor Kern- en Stralingsfysica, Celestijnenlaan 200D, B-3001 Leuven, Belgium \\
$^8$ Department of Physics and Astronomy and National Superconducting Cyclotron Laboratory, Michigan State University,\\ East Lansing, Michigan 48824-1321, USA \\
$^9$ Department of Theoretical Physics, Research School of Physical Sciences and Engineering, \\
     The Australian National University, Canberra ACT 0200, Australia \\
$^{10}$ Department of Physics, University of the Western Cape, Bellville  7530, South Africa \\}

\maketitle

\begin{abstract}
The nucleus $^{54}$Zn has been observed for the first time in an experiment at the SISSI/LISE3 facility of 
GANIL in the quasi-fragmentation of a $^{58}$Ni beam at 74.5~MeV/nucleon in a $^{nat}$Ni target. The 
fragments were analysed by means of the ALPHA-LISE3 separator and implanted in a 
silicon-strip detector where correlations in space and time between implantation and 
subsequent decay events allowed us to generate almost background free decay spectra for about 25 
different nuclei at the same time. Eight $^{54}$Zn implantation events were observed. 
From the correlated decay events, the half-life of $^{54}$Zn is determined to be 3.2$^{+1.8}_{-0.8}$~ms. 
Seven of the eight implantations are followed by two-proton emission with a decay energy of 1.48(2)~MeV. 
The decay energy and the partial half-life are 
compared to model predictions and allow for a test of these two-proton decay models.
\end{abstract}

\vspace*{0.2cm}
\vspace*{0.2cm}

Our understanding of nuclear structure is mainly based on results obtained with nuclei close 
to the line of stability. These studies allowed for understanding of the basic structure of the strong 
interaction which governs the interplay between neutrons and protons in an atomic nucleus. 
However, these nuclei close to stability cover only a very small range in isospin, i.e. their 
proton-to-neutron ratio is rather similar. With the advent of machines to produce radioactive 
nuclei, these basic concepts can now be tested with more and more exotic nuclei having a 
strong imbalance of neutrons and protons.

With these exotic nuclei being much further away from stability, also new phenomena 
appeared. For nuclei beyond the proton drip line, where the strong force can no longer bind all 
protons, one- and two-proton (2p) radioactivity was predicted more than 40 years ago by 
Goldanskii~\cite{goldanskii60}. For odd-Z nuclei, one-proton radioactivity was proposed to 
occur, whereas for medium- and heavy-mass even-Z nuclei the nuclear pairing energy renders one-proton emission 
impossible. In this case, two-proton emission is to be expected.

One-proton radioactivity was observed for the first time about 20 years ago by Hofmann et 
al.~\cite{hofmann82}.
Two-proton radioactivity was sought for many years without success. This 
research field experienced a strong boost with the advent of high-intensity 
projectile-fragmentation facilities. At these facilities, experimentalists could for the first time reach the 
most promising candidates for two-proton radioactivity. 
According to recent theoretical predictions, proton drip-line nuclei 
in the A=40-55 region were identified as the most promising 
candidates~\cite{brown02a,ormand97,cole96}. The recent observation of two-proton radioactivity 
of $^{45}$Fe~\cite{giovinazzo02,pfuetzner02} confirmed these predictions nicely. In other 
experiments, less promising candidates like $^{42}$Cr, and $^{49}$Ni could be shown to decay by 
$\beta$-delayed processes~\cite{giovinazzo01}. Beyond $^{45}$Fe, $^{48}$Ni and $^{54}$Zn were 
regarded as possible candidates to exhibit two-proton radioactivity. In the present letter, we 
report on the first observation of $^{54}$Zn and its decay by two-proton radioactivity.

$^{54}$Zn was produced by quasi-fragmentation reactions of  a primary $^{58}$Ni$^{26+}$ beam, accelerated to 
74.5~MeV/nucleon by the GANIL cyclotrons, which impinged with an average intensity  of  
4$\mu$A on a $^{nat}$Ni target of thickness 250mg/cm$^2$ installed in the SISSI device. 
The fragments were selected by the ALPHA-LISE3 separator which included a 50$\mu$m 
thick beryllium degrader in the intermediate focal plane of LISE. Two micro-channel 
plate (MCP) detectors at the first LISE focal plane and a detection set-up consisting 
of four silicon detectors installed at the end of the LISE3 beam line allowed us to identify the 
fragments on an event-by-event basis and to study their decay properties. 

The first silicon detector had a thickness of 300$\mu$m and served to measure the energy loss ($\Delta$E) 
of the fragments and their time-of-flight (TOF) with respect to the two MCP detectors and the 
radiofrequency (RF) of the cyclotrons. The second silicon detector, 300~$\mu$m thick, yielded
a second energy-loss signal, a TOF measurement with respect to the cyclotron RF, and served to 
detect $\beta$ particles from the radioactive decays in the adjacent double-sided silicon-strip 
detector (DSSSD). The DSSSD had a thickness of 500~$\mu$m and 16 strips (3~mm-wide) 
on each side. It was used to measure the residual energy of the fragments and their decay
characteristics. Finally, the fourth element was a 5~mm thick lithium-drifted silicon detector 
which served to detect $\beta$ particles from radioactive decays in the DSSSD. These detectors
yielded eight fragment identification parameters (two $\Delta$E signals, two residual energies
from both sides of the DSSSD, and four TOF measurements) which were used to unambigously
identify the different fragments and reject basically any background.

The experimental data were stored on tape on an event-by-event basis. To minimise the data 
acquisition dead time, we used two independent CAMAC/VXI branches. The trigger to start 
the event treatment switched from one system to the other after each event. Both branches 
were read-out via one VME branch. In order to avoid double triggering, 
we increased the trigger signal width to 20$\mu$s. This data acquisition 
system allowed us to treat two subsequent events as long as they are more than 20$\mu$s 
apart. The event treatment lasted about 300$\mu$s for each of the two branches, which means 
that we lost one event in the case where three events arrived within 300$\mu$s. However, 
these events still incremented a scaler which was read out event-by-event. Therefore, only 
events which followed a preceding event within less than 20$\mu$s got completely lost.

\begin{figure}
\begin{center}
\psfig{figure=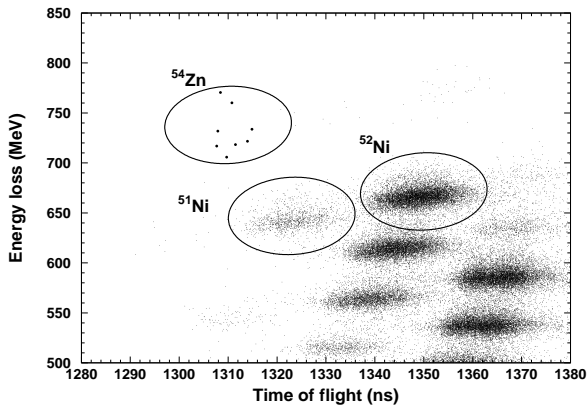,width=8.2cm}
\end{center}
\caption[]{Two-dimensional identification plot for the setting optimised on $^{54}$Zn. Plotted is 
                  the energy loss in the first silicon detector as a function of the time-of-flight 
                  between one micro-channel plate detector and the first silicon detector. Eight 
                  $^{54}$Zn events are identified. This figure presents only runs where a $^{54}$Zn nucleus was
                  observed.}
\label{fig:id}
\end{figure}

Figure~\ref{fig:id} shows the fragment identification matrix for the present experiment. To 
generate this plot, central values and widths for the distribution of each fragment on the eight 
identification parameters have been determined for isotopes with a high production rate and 
extrapolated for the very exotic nuclei. To be accepted all eight identification parameters of an 
event had to lie within 3 standard deviations of the predefined values. This procedure yields 
basically background free identification spectra. Thus, eight events have been attributed to $^{54}$Zn. 
This represents the first identification of this isotope. It is the most proton-rich zinc nucleus
ever observed and is predicted to be particle unstable with respect to two-proton 
emission by all modern mass predictions. However, the $Q_{2p}$ value varies widely (see 
below).

$^{54}$Zn has been produced by a two-proton pick-up reaction. From a 
production rate of about two nuclei per day, we estimate a production cross section of 
about $\sigma$~= 100~fb. 

\begin{figure}
\begin{center}
\psfig{figure=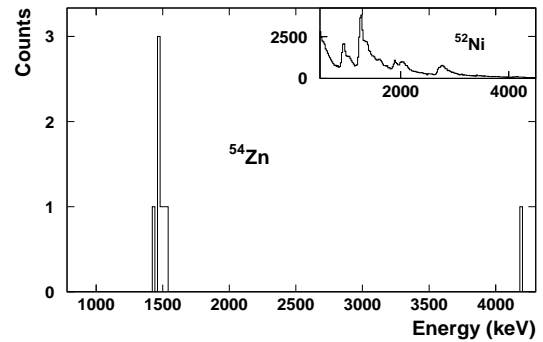,width=8.2cm}
\end{center}
\caption[]{Decay energy spectrum for the first decay events in the same pixel after an 
                  implantation of $^{54}$Zn. A decay energy of 1.48(2)~MeV is determined from the seven 
                  events in the peak. The eighth decay event has a decay energy of 4.19~MeV.
                  The insert shows the decay energy spectrum of $^{52}$Ni with the high-energy tails
                  of the protons groups due to $\beta$ pile-up.}
\label{fig:ep}
\end{figure}

As mentioned above, $^{54}$Zn is predicted to be two-proton unbound. Figure~\ref{fig:ep} shows 
the decay energy spectrum measured for the first decay event in the same x-y pixel after a $^{54}$Zn 
implantation. For eight implantations, we observe seven decay events with a decay energy of 
1.48(2)~MeV. None of these seven events has a coincident signal in the adjacent detectors, 
whereas the $\beta$-particle signals can be observed for neighboring 
nuclei which disintegrate by $\beta$ decay. The energy calibration was performed 
with well-known neighboring $\beta$-delayed proton emitters and $\alpha$ sources.
Both procedures need corrections: the $\beta$-delayed proton emitter calibration needs to be corrected
for the $\beta$ energy summing which was done with Monte Carlo simulations, whereas the $\alpha$-particle
calibration has to be corrected for the dead layer of the silicon detector. Both procedures
yielded consistent results. This calibration yields the total decay energy, which includes the
recoil of the daughter nucleus. It assumes that the pulse height defect is negligible.
The total error of the decay energy is a result of the error of the energy calibration and the
statistical error of the peak centroid.

The $\beta$ detection efficiency for the two detectors adjacent to the DSSSD was determined 
by means of $\beta$-delayed proton emitters like $^{52}$Ni. We determined a $\beta$ efficiency of 
40(5)\% for the Si(Li) detector and of 20(10)\% for the detector in front of the DSSSD. With 
these numbers, we determine a probability to miss all $\beta$ particles, if $^{54}$Zn would decay by a 
$\beta$-delayed mode with a 100\% branching ratio, of 0.16$^{+0.21}_{-0.15}$\%. 
Another indication that there is no $\beta$-particle emission 
comes from the fact that the full width at half maximum of the $^{54}$Zn peak in 
figure~\ref{fig:ep} is almost a factor of two narrower than e.g. the 1.3~MeV peak from 
$\beta$-delayed proton emission of $^{52}$Ni. This broadening for $^{52}$Ni comes from the energy 
loss of the $\beta$ particles in the DSSSD.

One decay event has an energy signal of 4.19~MeV. This event is in coincidence with a signal 
in the last detector identifying it as a $\beta$-delayed decay. We investigated the possibility 
that we missed the first decay event due to the data acquisition dead time. However, from the 
scaler content we found no evidence for such a loss. All events between the $^{54}$Zn implantation 
and the subsequent decay events ($^{54}$Zn and daughter decays) registered by the scalers are also 
on tape with the complete event information. This means that the only possibility to lose the 
first decay event for the eighth $^{54}$Zn is that this decay happened within 20$\mu$s of a 
preceding event. As the first observed decay event for this $^{54}$Zn implantation happens only 1.9~ms 
after the implantation event which has to be compared to the half-life of $^{54}$Zn (see below), we 
believe that it is highly likely that the 4.19~MeV event is indeed the first decay event. 
 
From these findings, we conclude that the branching ratio for 2p emission of $^{54}$Zn is BR~= 
87$^{+10}_{-17}$\%. We will use this branching ratio to determine the partial half-life for 2p decay of 
$^{54}$Zn which will be compared to model predictions below.

\begin{figure}
\begin{center}
\psfig{figure=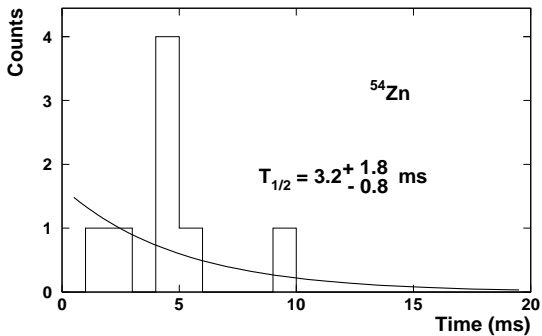,width=8.2cm}
\end{center}
\caption[]{The half-life of $^{54}$Zn is determined by means of the decay-time distribution
           of the first decay events after $^{54}$Zn implantation. The half-life determined is
           given in the figure.}
\label{fig:half}
\end{figure}

In figure~\ref{fig:half}, we present the decay time distribution for the eight decay events. 
We determine a half-life value of  3.2$^{+1.8}_{-0.8}$~ms. Together with the branching 
ratio for 2p emission determined above, we get a partial half-life for 2p emission of $^{54}$Zn of 
$T_{1/2}^{2p}$~= 3.7$^{+2.2}_{-1.0}$ms.

Another indication for the occurrence of 2p radioactivity is the observation of the decay of the 
2p daughter, $^{52}$Ni. This nucleus decays by $\beta$-delayed proton 
emission with proton energies of  1.06~MeV and 1.34~MeV and branching ratios of 4\% and 
10\%, respectively~\cite{dossat04}. We observe indeed two second decay events with energies in agreement 
with these expectations. However, $^{52}$Ni decays by $\beta$-delayed $\gamma$ emission with 
a branching ratio of 70\%. In these cases, we have a probability of only about 15\% to detect 
the $\beta$ particle in the DSSSD and therefore to be able to correlate it properly with the $^{54}$Zn 
implantation.

The experimental information collected above can only be explained consistently by the 
assumption of a decay by two-proton radioactivity of $^{54}$Zn with a branching ratio of 87$^{+10}_{-17}$\%.
All other possible decay modes do not yield a consistent picture. For $\beta$-delayed decays, 
we would expect to observe for a few decays the $\beta$ particle as well as a broadening of the 
peak. For $\alpha$ radioactivity, the barrier penetration half-life for an $\alpha$ particle with 
an energy of 1.48~MeV is several minutes, while the same barrier penetration lasts only about 
10$^{-17}$s in the case of one-proton emission with a decay energy of 1.48~MeV. We therefore 
conclude on the observation of ground-state two-proton radioactivity for $^{54}$Zn. In the 
following paragraphs, we will compare the experimental partial half-life of $T_{1/2}^{2p}$~= 
3.7$^{+2.2}_{-1.0}$ms and the 2p decay energy of $E_{2p}$~= 1.48(2)~MeV to different model predictions.

Starting from the 2p decay energy, we can determine the mass excess of $^{54}$Zn.
Using the mass excess of $^{52}$Ni as determined by means of the isobaric multiplet mass 
equation of $\Delta$m~= -22.64(4)~MeV~\cite{dossat04}, 
we obtain a mass excess of $\Delta$m~= -6.58(4)~MeV
for $^{54}$Zn. This mass excess compares well with e.g. the mass prediction of -6.34~MeV from 
J\"anecke and Masson~\cite{jaenecke88} and the mass extrapolation of -6.57(40)~MeV from 
Audi and co-workers~\cite{audi03}.

The decay energy $E_{2p}$ was recently predicted by several authors. Brown et 
al.~\cite{brown02a} proposed an energy of 1.33(14)~MeV. Ormand~\cite{ormand97} determined an 
energy of  1.97(24)~MeV, whereas Cole~\cite{cole96} predicted a value of 1.79(12)~MeV. All these 
theoretical decay energies are reasonably close to our experimental datum. However, when used in a 
simple barrier-penetration di-proton model, which is expected to yield only a lower limit for 
the partial half-life for 2p decay~\cite{grigorenko01}, we obtain values between 20~ms and 
10$^{-7}$~ms, a rather large spread.

\begin{figure}
\begin{center}
\psfig{figure=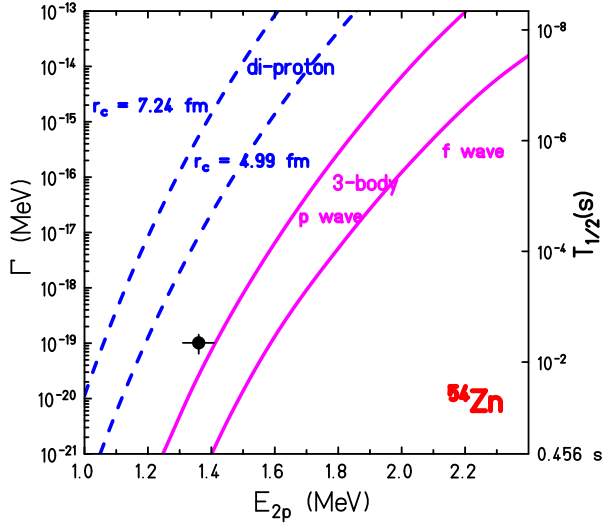,width=8.2cm,angle=-90}
\end{center}
\vspace*{-0.6cm}
\caption[]{Relation between half-life / decay width and decay energy as determined by the three-body
                 model of Grigorenko et al.~\cite{grigorenko03}. The picture compares the 
                 experimental datum to the three-body model (solid lines) with two different assumptions for 
                 the proton emitting orbital and with the di-proton model (dashed lines) for two different channel 
                 radii.}
\label{fig:grigo}
\end{figure}

Such a di-proton model completely neglects any nuclear structure or any dynamics of the 
decay. Grigorenko and co-workers~\cite{grigorenko01,grigorenko03} 
developed a model which explicitly includes the decay 
dynamics by modelling the proton-proton and proton-core interaction in the decay. Their 
three-body model result is compared to our experimental result in figure~\ref{fig:grigo}. We obtain 
reasonable agreement between experiment and theory, if we assume that the two protons are 
emitted from a $p$ orbital, as expected from simple shell-model arguments. The three-body 
model yields much longer half-lives for a given decay energy than the simple di-proton 
model.

The three-body model contains only little nuclear structure information, but treats the 
dynamics in a reasonable fashion. Brown and Barker developed the R-matrix approach to 
apply it for 2p emission~\cite{barker01,brown03}. This model includes the $s$-wave proton-proton 
interaction as an intermediate state. The nuclear structure input is the spectroscopic 
factor \cite{anyas-weiss74} $S = \left( \frac{A}{A-2} \right)^{\lambda } \; G^{2}(pf) \; C(A,Z)  $
where $  G^{2} = 5/16  $, $\lambda$=6, $  A  $ is the mass of the parent 
nucleus ($A~= 54$), and $C(A,Z) =\, \mid <\Psi (A-2,Z-2)\mid \psi _{c}\mid \Psi (A,Z)>\mid ^{2}  $
is the cluster overlap for the di-proton cluster wavefunction $\psi_{c}$ in
the $pf$ shell with  $L=0, S=0$ and $T=1$ in the SU3 basis. The $pf$-shell wavefunctions $\Psi$ 
were obtained with the recent GPFX1 interaction~\cite{honma04} with the result $C~= 0.39.$
For the di-proton~- nucleus potential, we take a Woods-Saxon form plus a uniform-sphere Coulomb potential 
with radius $R_{C}~= r_{C} A^{1/3}$. The Woods-Saxon parameters are
$R~= r_{0} A^{1/3}$ for the radius, $a_{0}$ for the diffuseness, and a well depth adjusted
to reproduce the resonance energy. The potential parameters are taken
from an analysis of low-energy deuteron scattering \cite{daehnick80}:
$r_{0}~= 1.17$~fm, $a_{0}~= 0.72$~fm, $r_{C}~= 1.30$~fm. 
These are the same potential parameters used for the calculation of the  
di-proton decay for $^{45}$Fe in reference~\cite{brown03}.

With the experimental $Q$ value of 1.48(2) MeV,
the resulting half-life is 16$^{+10}_{-6}$ ms, whereas the value is 9$^{+5}_{-3}$~ns 
when the $p-p$ resonance is ignored. The experimental decay rate
is 2-10 times faster than the theoretical value which takes into account 
the $p-p$ resonance. A similar enhancement
was found in the comparison of experiment and theory for $^{45}$Fe~\cite{brown03}.
Pairing correlations due to configuration mixing beyond the $pf$ shell may account for this 
enhancement. In reference~\cite{decowski78}, an enhancement factor of about two (the ratio of
the $\epsilon$ factors in columns 3 and 5 of Table 3 of this paper) due to mixing 
with the $sd$ and $g_{9/2}$ orbitals was calculated for the two-neutron transfer
in the $(p,t)$ reaction in the $pf$ shell. Calculations of the
two-proton decay that include orbitals outside the $pf$ shell
remain to be carried out.

In summary, we observed for the first time the new isotope $^{54}$Zn. As predicted by modern 
mass models, it decays by two-proton emission with a decay energy of $E_{2p}$~= 
1.48(2)~MeV. The two-proton branching ratio is 87$^{+10}_{-17}$\% and the total half-life  
was determined to be 3.2$^{+1.8}_{-0.8}$~ms yielding  a 2p partial half-life of 
$T_{1/2}^{2p}$~= 3.7$^{+2.2}_{-1.0}$ms. The two advanced models able to describe two-proton 
radioactivity achieve reasonable agreement with our experimental data. Future studies will try 
to find more cases of two-proton radioactivity and investigate the decay in more detail by measuring the
p-p angular correlation and the energy of the individual protons. 

We would like to acknowledge the continous effort of the whole GANIL staff for ensuring 
a smooth running of the experiment. This work was supported in part by the US
National Science Foundation under grant number PHY-0244453 and the South African NRF ISL grant GUN 
No. 2068517, as well as by the Conseil R\'egional 
d'Aquitaine.

\end{document}